\documentclass[11pt,a4paper]{article} 
\pdfoutput=1
\usepackage{jheppub}

\pdfoutput=1

\usepackage{xspace}
\usepackage{epsfig}
\usepackage{amssymb,amsmath}
\usepackage{multicol}
\usepackage{multirow}
\usepackage{url}

\def\be{\begin{equation}}
\def\ee{\end{equation}}
\def\bea{\begin{eqnarray}}
\def\eea{\end{eqnarray}}
\def\beal{\begin{align}}
\def\eeal{\end{align}}

\def\hc{\ {\rm h.c}}
\def\nn{\nonumber}

\def\be{\begin{equation}}
\def\ee{\end{equation}}
\def\bea{\begin{eqnarray}}
\def\eea{\end{eqnarray}}
\def\ca{c_{\alpha}}
\def\sa{s_{\alpha}}
\def\gprime{g^{\prime}{}}
\def\sw{s_{W}}
\def\cw{c_{W}}
\def\c2w{c_{W}^2}
\def\s2w{s_{W}^2}
\def\cchi{c_{\chi}}
\def\schi{s_{\chi}}

\def\mplus{m_{2}}
\def\mm{m_{1}}

\def\mzp{m_{Z'}}
\def\epsh{\epsilon_H}
\def\gD{g_D}

\def\lamd{\lambda_D}

\def\Zprime{Z'}

\title{Dark Higgs Models at the 7 TeV LHC} 
\author[a]{Erich Weihs}
\author[a]{and Jos\'e Zurita}

\affiliation[a]{Institut f{\"u}r Theoretische Physik, Universit\"at Z\"urich, Winterthurerstrasse 190, \\ CH-8057 Z\"urich, Switzerland.}

\emailAdd{erich.weihs@physik.uzh.ch}
\emailAdd{jzurita@physik.uzh.ch}

\abstract{
We study how collider data and electroweak precision observables affect the parameter space of models including a new \emph{dark} force mediated by a massive U(1) gauge boson. It acquires mass via a Higgs mechanism in the dark sector which is connected to the Standard Model through kinetic mixing of the two U(1) gauge bosons and the Higgs potential. 

We assess the impact of the 7 TeV LHC and show that most of the parameter space of the model can be probed with an integrated luminosity of $15~\textrm{fb}^{-1}$.
}

\arxivnumber{1110.5909}
\begin{document}
\begin{flushright}
JHEP 1202 (2012) 041\\
ZU-TH 20/11\\
LPN11-56
\end{flushright}
\maketitle

\section{Introduction}
\label{sec:intro}

Extensions of the SM often include one or more additional abelian gauge symmetries beside the Standard Model $U(1)_Y$ hypercharge. Compelling motivation for these extensions arises from grand unified theories (GUT) and from string theory. The phenomenology of new abelian gauge groups has been widely studied in the literature
\cite{Langacker:2008yv,Appelquist:2002mw,Rizzo:2006nw,Dienes:1996zr,Leike:1998wr,Hewett:1988xc,Erler:2011iw,Fox:2011qd,delAguila:2010mx,Williams:2011qb,Babu:1997st}.

Moreover, in the recent years there was a renewed interest in this kind of models, based on the key observation \cite{ArkaniHamed:2008qn,Pospelov:2008jd} that an extra $U(1)_D$ GeV gauge boson would be able to naturally explain the anomalies observed in indirect dark matter detection experiments, like the reported PAMELA result on the positron fraction \cite{Adriani:2008zr}.
The extra gauge symmetry may be hidden from the SM particles, which are singlets under the new force: they constitute the \emph{visible} sector. The particles charged under the new $U(1)_D$ and singlets under the SM gauge groups are often referred to as the \emph{dark} (or hidden, or secluded) sector.

The connection between the \emph{dark} and \emph{visible} sectors is established through mixing operators. One candidate term is kinetic mixing of $U(1)_Y$ with $U(1)_D$ \cite{Foot:1991bp,Holdom:1985ag}. Since cosmological considerations (like Big Bang Nucleosynthesis) severely constrain a massless gauge boson, the extra $U(1)$ symmetry has to be broken. Its breakdown  can be achieved through the introduction of a new Higgs boson, $h_D$, which can naturally mix with the SM Higgs \cite{Schabinger:2005ei,Patt:2006fw,Wells:2008xg,Chang:2008cw,Barbieri:2005ri}, thus providing an extra \emph{portal} between the two worlds. These extensions of the SM have also been studied in the context of electroweak phase transition (EWPT) \cite{Espinosa:1993bs,Espinosa:2007qk,Choi:1993cv} and dark matter (DM) since the dark sector provides natural DM candidates \cite{Silveira:1985rk,McDonald:1993ex,Davoudiasl:2004be,Barger:2010yn,Burgess:2000yq,Mambrini:2011ik,Gopalakrishna:2009yz,Cai:2011kb,Boehm:2003hm,Barbieri:2006dq,Chang:2007ki,Feldman:2007wj,Pospelov:2008zw,Pospelov:2007mp,Liu:2009pt,Chun:2010ve,Andreas:2008xy,Tytgat:2010bt,Arina:2010rb,Barger:2010mc,Batell:2010bp}. 

While the extra $U(1)$ models are very constrained from current experimental data \cite{:2009pw,Feng:2009hw,Abazajian:2010sq,Hook:2010tw,Mambrini:2011dw}, the non-observation of a Higgs boson yields very mild bounds on the Higgs portal parameters at present. In this paper we explore the constraints and detectability prospects of the Higgs sector at colliders. Very recent work \cite{Englert:2011yb} was focused in the potential signatures at the LHC for large luminosities (${\cal O} (30)$ fb$^{-1}$, see also Refs.\cite{Barger:2007im,Barger:2008jx,Bowen:2007ia,Cerdeno:2006ha,  O'Connell:2006wi,BahatTreidel:2006kx,Gopalakrishna:2008dv,Bock:2010nz} for older studies).
Our interest resides in the reach of the early LHC data ($\sqrt{s}=7$ TeV, with a total integrated luminosity less or equal than 15 fb$^{-1}$). Similar work was already done in the context of the MSSM in ref~\cite{Carena:2011fc}.

Due to the mixing with hypercharge, the dark gauge boson can be in conflict with electroweak precision data, such as the Z mass or the effective weak mixing angle. Thus the most natural options, already considered in the literature, is to have either a very heavy (TeV scale) $\Zprime$~\cite{Kumar:2006gm,Grossmann:2010wm,Chang:2006fp}
 or very light (GeV) boson~\cite{Gondolo:2011eq,Cheung:2009qd,Ahlers:2008qc,Kang:2010mh}. The latter scenario is well-motivated when looking to find a unified explanation of recent results of DM, as suggested in ref~\cite{ArkaniHamed:2008qn}. In this work we take an agnostic attitude and consider the $\Zprime$ mass a free parameter. 

This paper is organized as follows: in Section \ref{sec:model} we review the model under consideration. In Section \ref{sec:scans} we explain in detail the scan of the parameter space, all the constraints under consideration and the LHC expected reach for different scenarios. Section~\ref{sec:results} contains the numerical results of our analysis. Finally, we conclude in Section~\ref{sec:conclu}.  
%
\section{Model review}
\label{sec:model}

Generic dark sector models were discussed in detail in the literature (e.g. \cite{Wells:2008xg}). 
In this section we briefly review the model used in our study. The Lagrangian can be
 written as follows, 
\be
{\cal L} = {\cal L}_{SM}+{\cal L}_{Dark}+{\cal L}_{mix}  \, ,
\ee 
where we have split the contribution into the SM-piece, the dark sector and the mixing between the two sectors.
For the dark sector, we would like to add the minimum field content. Thus, we include a new dark gauge boson $X$ and a dark Higgs field $H_D$. The dark sector might contain fermions, which are SM singlets and charged under $U(1)_D$. These fermions are, however, irrelevant in the present context. The dark Higgs field will give mass to the $X$ boson after spontaneous breakdown of the gauge symmetry.  We pick a $U(1)$ gauge group for simplicity; that is not to say that the dark sector has to be that simple, but that we choose to parametrize it in a simple way.
It is clear that many other, richer possibilities (from a phenomenological point of view) can also be considered\footnote{One could argue that the details of the dark sector at energies above LEP and SLC could be absorbed into the low energy ${~\rm GeV}$ scale parameters by integrating out the heavy sector. Another option is to work with a different dark gauge group. We will stick, for the sake of simplicity, to this minimum extra added field content.    }.
Under these assumptions, the dark Lagrangian reads
\be\label{eq:LagDark}
{\cal L}_{Dark} = (D_\mu H_D)^{\dagger} (D_\mu H_D) + \mu_D H_D^{\dagger} H_D - \lambda_D (H_D H_D^{\dagger})^2 - \frac{1}{4} X_{\mu \nu} X^{\mu \nu} + \dots
\ee
where $D_\mu= \partial_{\mu} + i \gprime Y B_{\mu}  + i g T^{a} W_{\mu}^a+ i g_D Q_D X_{\mu} $ is the covariant derivative, $g_D$ the dark $U(1)_D$ gauge coupling, $X_{\mu \nu}$ its gauge strength tensor and $Q_{D}$ is the charge under the dark force. The last term is the kinetic term for the dark gauge field, while the remaining terms correspond to the kinetic term for the complex scalar Higgs and the dark Higgs potential. The ellipsis stands for other terms not relevant for our study. 
The \emph{mixed} Lagrangian will depend upon how $X$ and $H_D$ couple to the SM. In our setup, it is natural to consider kinetic mixing between $X_{\mu}$ and $B_{\mu}$ 
and a mixing term in the Higgs potential \cite{Schabinger:2005ei,Patt:2006fw}, since these two are the only renormalizable operators relating $X$ and $H_D$ to the SM.\footnote{As noted in ref~\cite{Gopalakrishna:2008dv}, there are other such operators if the dark fermions are also taken into account. 
} With these assumptions we have
\be\label{eq:lagmix}
{\cal L}_{mix}=\frac{\epsilon_A}{2} B_{\mu \nu} X^{\mu \nu} + \epsilon_H (H H^{\dagger}) (H_D H_D^{\dagger}),
\ee
where $B_{\mu \nu}$ is the $U(1)_Y$ hypercharge field strength tensor and $H$ is the SM Higgs doublet.
It is well known that $\epsilon_A$ has to be small in order to be compatible with current experimental limits (see \cite{Hook:2010tw} and references therein). The constraints on $\epsilon_H$ are less stringent, given our current knowledge of the Higgs sector. 

For the sake of completeness, we write down the SM Lagrangian,
\begin{eqnarray}\label{eq:LagSM}
{\cal L}_{SM}&=& (D_\mu H)^{\dagger} (D_\mu H) + \mu H H^{\dagger} - \lambda (H H^{\dagger})^2  \nonumber \\
&+& \sum_{f} y_f ( \bar{f}_L H f_R + \hc )  - \frac{1}{4} (B_{\mu \nu} B^{\mu \nu} + W^{\mu}_a W^{a}_{\mu})+ \dots,
\end{eqnarray}
where $y_f$ is the Yukawa coupling for the SM fermion $f$ and the ellipsis indicates the presence of other terms not relevant for our study.
\subsection{Gauge sector}
In order to derive the interactions in the mass eigenstate basis, we have to proceed in several steps. First, one has to diagonalize the kinetic terms for the gauge bosons. This can be achieved by performing a field redefinition of $B_{\mu}$ and $X_{\mu}$. After this, one finds that the covariant derivative has changed in such a way that now the dark sector interacts directly with the $B_{\mu}$. Since
we want the $U(1)_D$ gauge group to be broken, the vacuum expectation value of $H_D$ will contribute to the masses of the $Z$ and the $\Zprime$, while the photon will remain massless.

The Lagrangian involving both $U(1)$ strength tensors is given by
\be
{\cal L} = - \frac{1}{4} \bigl( B_{\mu \nu}   B^{\mu \nu}  + X_{\mu \nu}   X^{\mu \nu} - 2 \epsilon_A  B^{\mu \nu}  X_{\mu \nu} \bigr).
\ee 
In order to diagonalize the kinetic term, we perform the following redefinition of the fields \cite{Babu:1997st} first:
\be\label{eq:fieldredef}
B_{\mu} \to B_{\mu} + \frac{\epsilon_A}{\sqrt{1-\epsilon_A^2}} X_{\mu} \, , \qquad \, X_{\mu} \to  \frac{1}{\sqrt{1-\epsilon_A^2}} X_{\mu}   \, .
\ee
Then the covariant derivative reads\footnote{In our convention, the SM Higgs doublet has a Y=+1/2, and the dark Higgs doublet also has $Q_D=+1/2$.}
\be
D_{\mu} = \partial_{\mu} + i \gprime Y B_{\mu} + i g T^3 W_{\mu}^3 + i \left( g_D \frac{Q_D}{\sqrt{1-\epsilon_A^2}} + \gprime \frac{\epsilon_A Y}{\sqrt{1-\epsilon_A^2}} \right) X_{\mu} \, ,
\ee
and the mass matrix of the neutral gauge bosons becomes 
\be
m_{Z_0}^2
\Biggl(
\begin{array}{ccc}
\s2w & - \cw \sw & a \s2w \\ 
-\cw \sw & \c2w & - a \cw \sw\\ 
a \s2w & -a \cw \sw& a^2 \s2w + \Delta 
\end{array}
\Biggr) 
\, ,
\ee
where $s_W, c_W$ are the sine and cosine of the usual SM electroweak mixing angle and
\be 
m^2_{Z_0}=(g^2+\gprime^2) \frac{v^2}{4} , \qquad \, 
m^2_{X_0}=g_D^2  \frac{v_D^2}{4 (1-\epsilon_A^2)} , \qquad
\Delta=\frac{m_{X_0}^2}{m_{Z_0}^2},  \qquad 
a=\frac{\epsilon_A}{\sqrt{1-\epsilon_A^2}} \, .
\ee
One of the mass eigenvalues is zero, corresponding to the photon eigenstate, and the two others are given by
\be\label{eq:gaugebosonmasses}
M^2=\frac{m_{Z_0}^2 }{2} \bigl[ (1 + s_W^2 a^2 + \Delta ) \pm \sqrt{(1 + s_W^2 a^2 + \Delta )^2-4 \Delta} \bigr] \, .
\ee
Due to the smallness of  $\epsilon_A$ it is well justified to take the gauge boson masses at their tree level values, namely, to assume $m_{Z}=m_{Z_0}$ and $m_{\Zprime}=m_{X_0}$. We have numerically checked that this approximation has an error below 0.02 \%.
The relation between mass and interaction eigenstates is given by
\be
\Biggl(
\begin{array}{c}
B_{\mu} \\
W^3_{\mu} \\
X_{\mu}
\end{array}
\Biggr) =
\Biggl(
\begin{array}{ccc} 
c_W & -s_W \cchi & s_W \schi \\
s_W & c_W \cchi & -c_W \schi \\
0 & \schi & \cchi 
\end{array}
\Biggr) 
\Biggl(
\begin{array}{c}
A_{\mu} \\
Z_{\mu} \\
\Zprime_{\mu}
\end{array}
\Biggr) \, ,
\ee
and the new gauge boson mixing angle by
\be
\tan 2 \chi = \frac{-2 s_W a}{1-s_W^2a^2-\Delta} \, .
\ee
\subsection{Higgs sector}
\label{subsec:minHiggspot}
In the unitary gauge, one has
\begin{eqnarray}
H= \frac{1}{\sqrt{2}} \left( \begin{array}{c} 0 \\ v + h  \end{array} \right) \,, \qquad \qquad H_D= \frac{1}{\sqrt{2}} (v_D + h_D) \, ,
\end{eqnarray}
and the minimization of the Higgs potential yields
\be\label{eq:min}
\mu= \lambda v^2 - \epsilon_H \frac{v_D^2}{2} \, , \qquad \qquad \mu_D= \lambda_D v_D^2 - \epsilon_H \frac{v^2}{2} \, . \ee
The squared mass matrix of the Higgs sector reads
\be
{\cal M}^2 = 
\left(
\begin{array}{cc}
 2 \lambda v^2 & -\epsilon_H v v_D    \\
 -\epsilon_H v v_D   & 2 \lambda_D v_D^2    \, 
\end{array}
\right)\, ,
\ee
with its eigenvalues given by
\be
m^2_{1,2} =\lambda v^2 + \lambda_D v_D^2  \mp \sqrt{( \lambda v^2 - \lambda_D v_D^2)^2 + \epsilon_H^2 v^2 v_D^2} \, ,
\ee
where $m_2 > m_1$. The mass eigenstates read
\be\label{eq:hmix}
h_2 = \ca h - \sa h_D \, , \qquad \qquad h_1 =  \sa h + \ca h_D \, ,
\ee
and the mixing angle is given by 
\be
s_{2 \alpha}= \frac{\epsilon_H v_D v}{\sqrt{( \lambda v^2 - \lambda_D v_D^2)^2 + \epsilon_H^2 v^2 v_D^2}} \, , \qquad 
c_{2 \alpha}= \frac{ \lambda v^2 - \lambda_D v_D^2}{\sqrt{( \lambda v^2 - \lambda_D v_D^2)^2 + \epsilon_H^2 v^2 v_D^2} }\, .
\ee
We define the effective Higgs coupling as the coupling in our model normalized to the SM case. Using  eq.~(\ref{eq:hmix}) in eqs.~(\ref{eq:LagSM}) and ~(\ref{eq:LagDark}), one has
\be\label{eq:effHcoup}
g_{h_1 WW} = g_{h_1f\bar{f}}=\sa \, , \qquad g_{h_2 WW} = g_{h_2 f \bar{f}} = \ca \, .
\ee
The couplings to $Z-\Zprime$ read
\be\label{eq:hZZ}
g_{h_2 Z_1 Z_2} = \ca g_{h Z_1 Z_2} - \sa g_{h_D Z_1 Z_2} \Delta \frac{v}{v_D} \, , \qquad g_{h_1 Z_1 Z_2} =  \sa g_{h Z_1 Z_2} + \ca g_{h_D Z_1 Z_2} \Delta \frac{v}{v_D} \, ,
\ee
where $Z_{1,2}=Z,\Zprime$, the $g_{H Z_1 Z_2}$ factors are given in table \ref{tab:hZZ}.
Due to the smallness of the kinetic mixing one finds that $g_{hZZ} \approx g_{h_D \Zprime \Zprime} \approx 1$, while all the other 
are at least suppressed by a power of $\epsilon_A < 0.03$. Therefore, one has that the coupling of $h_1$ ($h_2$) to the SM particles is suppressed by a factor of $\ca~( \sa)$ with respect to the values of the SM Higgs.
\begin{table}[htdp]
\begin{center}
\begin{tabular}{|c|c|c|c|c|}
\hline
H & $ZZ$ & $\Zprime \Zprime$ & $Z \Zprime$ \\
\hline
$h$ &  $(-\cchi + a \schi \sw)^2 $&  $(\schi + a \cchi \sw)^2$ & $ (-\cchi + a \schi \sw)  (\schi + a \cchi \sw)$ \\
\hline
$h_d$ & $\schi^2$ & $\cchi^2$ & $\schi \cchi$  \\
\hline
\end{tabular}
\end{center}
\caption{$g_{H Z_1 Z_2}$ couplings.}
\label{tab:hZZ}
\end{table}%

There are also interactions involving three and four Higgs fields, as well as two gauge bosons plus two Higgs fields. These decay modes constitute what we will call, from now on, \emph{non-standard} ($\text{Non-SM}$) Higgs decay modes, namely, those that do not appear when considering the SM Higgs boson. They could be important, for instance, if there is a significant fraction in the $h_2 \to h_1 h_1$ or $h_2 \to Z^{\prime} Z^{\prime}$  at LEP, like in the buried Higgs scenario \cite{Bellazzini:2009xt}.
In our setup we assumed that the decay width of the $Z'$ into Standard Model particles is negligible, since its couplings to Standard Model particles are suppressed by a factor of $\epsilon_A$ with respect to the couplings of the $Z$.
The decay width of a Higgs boson into two gauge bosons $Z_1$ and $Z_2$ is given by
\bea
\Gamma (H \to Z_1 Z_2) = \frac{g^2 m_H^3  g^2_{H Z_1 Z_2} S}{64 \pi m_W^2} && \frac{m_Z^4}{m^2_{Z_1} m^2_{Z_2}}  \left[ 1 - \frac{(x_1+x_2)}{2}  + \left(\frac{x_1-x_2}{4}\right)^2 \right]^{1/2}  \times \nn \\ && \left[ 1+ \frac{5}{8} x_1 x_2 + \frac{x_1^2+x_2^2}{16} - \left(\frac{x_1+x_2}{2}\right) \right]  \, ,
\eea
where $H=h_1, h_2$,  $x_{1,2}=(2 m_{Z_{1,2}}/m_H)^2$, $g_{H Z_1 Z_2}$ can be read from table \ref{tab:hZZ} and $S$ is a symmetry factor, 1/2 if $Z_1=Z_2$, 1 otherwise.
  The partial widths of the heavy Higgs boson into light ones is
   \be\label{eq:hhdhd}
 \Gamma (h_2 \to h_1 h_1 ) = \frac{1}{32 \pi m_{h_2}} \sqrt{1-\frac{4 m^2_{ h_1} }{m^2_{h_2}}} \lvert g_{h_2 h_1 h_1} \rvert^2 \, ,
 \ee
 where the trilinear Higgs coupling $g_{h_2 h_1 h_1}$ is given by
 \be
  g_{h_2 h_1 h_1} = 2 \bigl\{ 3 \sa \ca \bigl(\lambda v \sa -\lambda_D v_D \ca \bigr)
   - \frac{\epsilon_H}{4}  \bigl[ v \ca (3 c_{2 \alpha} - 1) + v_D \sa  (3 c_{2 \alpha} + 1)    \bigr]  \bigr\} \, .
   \ee
Due to the rescaling of the Higgs-to-Standard Model couplings the Higgs production cross sections are suppressed by a factor of $\sa^2$ for $h_1$ ($\ca^2$ for $h_2$). Consequently, in the case where one can neglect the $\text{non-SM}$ decays, there is always one Higgs boson whose rate is suppressed at most by a factor of 1/2. The branching fractions into SM particles will be suppressed by a factor of $1-{\rm Br } (h_i \to \text{non-SM})$. Therefore, the total rate for any Higgs boson into SM particles is always lower than in the SM by a factor of
\be\label{eq:ratesup}
g_{h_i WW}^2 \left(1-{\rm Br} (h_i \to \text{non-SM}) \right)  \, .
\ee

\section{Numerical analysis: parameter scans and constraints}
\label{sec:scans}
%
\subsection{Parameter scans and pre-LHC constraints}
To explore the parameter space of the model a random parameter scan was performed using the \textsc{Cuba}-library \cite{Hahn:2004fe}. We chose as input parameters the physical parameters $\mm$, $\mplus$, the mixing angle $\alpha$, $\gD$, $\mzp$ and the kinetic mixing parameter $\epsilon_A$ with values in the ranges according to table \ref{tab:parametertable}.
\begin{table}[htdp]
\begin{center}
  \begin{tabular}{| c | c | c | c | c | c | }
    \hline
    $\mm~ [\rm{GeV}] $ & $\mplus~ [\rm{GeV}]$ & $\alpha$ & $\mzp~[\rm{GeV}]$ & $\gD$ & $\epsilon_A$ \\ \hline
    $[1 ; 400]$ & $[1 ; 600]$ & $[0 ; \pi]$ & $[0 ; 1000]$ & $[0 ; 1]$ & $[0 ; 0.3]$ \\ \hline
  \end{tabular}
\end{center}
\caption{Ranges of the parameter scan.}
\label{tab:parametertable}
\end{table}
We focused on Higgs masses below 600 GeV since the LHC experiments have published exclusions in that mass range and the phenomenology of a heavy singlet Higgs has been studied elsewhere (see, for instance, ref~\cite{Bowen:2007ia}).

The potential parameters were computed using 
 \be\label{eq:lamSM}
 \lambda= \frac{1}{4 v^2} \bigl[ m^2_1 
 \bigl( 1 -c_{2 \alpha}  \bigr) + m^2_2 \bigl( 1 +c_{2 \alpha}  \bigr) \bigr] \, ,
 \ee
  \be
 \lambda_D= \frac{1}{4 v_D^2} \bigl[ m^2_1
\bigl( 1 +c_{2 \alpha}  \bigr)
 + m^2_2
\bigl( 1 -c_{2 \alpha}  \bigr)
 \bigr] \, ,
 \ee
 \be
 \epsilon_H = \frac{1}{2 v v_D}  (m^2_2 -  m^2_1) \, s_{2 \alpha}  \, .
 \ee
We also required the points to respect the positivity conditions, eq.~(\ref{eq:min}), thereby ensuring the proper minimalization of the potential. 
Motivated by the discussion of the electroweak phase transition in similar models (see, for example, \cite{Profumo:2007wc} and references therein), we discarded points with nonperturbative potential parameters by requiring $\epsh \leqslant 0.5$ and $\lambda,\, \lamd \leqslant 1$. This also limits the contribution of the invisible decay modes to the total width of the Higgs bosons such that their values stay within the validity of the narrow width approximation (i.e. $\Gamma^{\text{tot}}_i / m_i < .05$), which is required in order to interpret the exclusion limits set by collider data on the rates of the Higgs boson as the product of the production cross-section times branching ratio in a particular channel.

Constraints from direct searches were applied using \texttt{HiggsBounds 2.1.1}  \cite{Bechtle:2008jh,
Bechtle:2011sb}, where points are excluded at the 95\% confidence level. In the low-mass region (below 114.4 GeV) the main exclusion channels are the LEP searches for a Standard Model-like Higgs \cite{Barate:2003sz,Abbiendi:2002qp,Schael:2006cr} and a Higgs-like scalar decaying completely invisibly \cite{:2001xz,Achard:2004cf,:2007jk}. In some cases the decay $h_2\rightarrow h_1 h_1 \rightarrow 4 b\; \text{or}\; 4 \tau$ was also constrained directly by the corresponding LEP MSSM searches \cite{Schael:2006cr}. In the high-mass region (120-200 GeV) the Tevatron searches  were also used to bound the parameter space \cite{:2010ar,Aaltonen:2010sv}.

Electroweak precision data also limit the parameter space of our model in a significant way \cite{Chanowitz:2008ix,Peskin:2001rw}. In order to assess the effect of a complete parameter fit, we used model independent bounds on the kinetic $Z\;-Z'$ - mixing \cite{Hook:2010tw} to constrain $\epsilon_A$ and computed the contribution of the extended gauge and Higgs sectors to the Peskin-Takeuchi $S$ and $T$ parameters \cite{Peskin:1990zt} using FormCalc \cite{Hahn:2009bf}. For the two Higgs bosons $h_1$ and $h_2$, it is given by 
\be 
S = \ca^2 S^{\text{SM}} (m_1) + \sa^2 S^{\text{SM}} (m_2),
\ee
where $S^{\text{SM}}$ denotes the contribution of a Standard Model Higgs with respect to the reference mass $m_h=120$ GeV (and analogously for the T-parameter). 
The tree-level contribution of the $Z'$ to the oblique parameters is \cite{Holdom:1990xp}
\be \label{eq:SZprime}
\alpha_{EW} S = 4 \cw^2 \sw^2 \frac{\cw^2 - \Delta}{(\Delta-1)^2} \epsilon_A^2,  \quad \text{and}\quad \alpha_{EW} T = - \sw^2 \frac{\Delta}{(\Delta-1)^2} \epsilon_A^2, 
\ee
which diverge as $m_{Z'} \rightarrow m_Z$. We are however confident that the formulae are valid as long as $|\frac{\sw \epsilon_A}{1-\Delta}| \ll 1$. Due to the constraints on $\epsilon_A$ that we implemented this condition is always fulfilled in our scan. 
Since we study the Higgs sector of this theory at the LHC, we were interested in how a $Z'$ with suitably chosen properties can relax the upper mass limit on the Standard Model Higgs mass from the S and T parameter fit, which is the case when the tree level contributions are enhanced through $m_{Z'} \rightarrow m_Z$. We neglect loop contributions of the $Z'$ via the ordinary photon and $W$,$Z$ gauge boson self energies, since their size would be of order of the Standard Model gauge sector contributions to the neutral current amplitude, but suppressed by an additional factor of $\epsilon_A^2$. In the threshold region around $m_{Z'}=m_Z$ it is suppressed even further by the strong constraints on $\epsilon_A$. We also neglected dark fermions, since their contribution would only enter the forementioned at the two-loop level. We set $U=0$ and required a parameter space point to lie inside the $2\sigma$ contour in the $S-T$-plane provided by the Gfitter collaboration \cite{Flacher:2008zq,gfitterURL,Bardin:1999yd,Arbuzov:2005ma}.

%
\subsection{LHC data and projections}
\label{sec:LHC}
%
\begin{table}[t]
\begin{center}
\begin{tabular}{|c|cc|c|c|c|cl}
\hline
\rule{0mm}{5mm}
\multirow{2}{*}{Channel} &
\multicolumn{2}{|c|}{Lum. (fb$^{-1}$)} &
\multirow{2}{*}{What we do} &
Mass range &
\multirow{2}{*}{Ref.} \\
 & ATLAS & CMS  & & (GeV) &
 \\ [0.3em]
\hline
\rule{0mm}{5mm}
$p p \to H \to WW$ & 1.7  &   1.5 & Comb. & 115-600 & \cite{ATLAS:htoWW2l2nu,Collaboration:2011as,CMS:hWW}
\\ [0.3em]
\hline
\rule{0mm}{5mm}
$p p \to H \to ZZ$ & $1.04$--$2.28$ & $1.1$--$1.7$ & Comb. & 120-600 &   \cite{Aad:2011uq,Collaboration:2011va,Aad:2011ec,CMS:htoZZto4l,CMS:htoZZto2l2nu,CMS:htoZZto2l2j,CMS:htoZZto2l2taus}
\\ [0.3em]
\hline
\rule{0mm}{5mm}
$p p \to H \to \gamma \gamma$  & 1.08 & 1.7 & Comb. & 110-150 &  \cite{Collaboration:2011ww,CMS:Htogaganote}
\\ [0.3em]
\hline
\rule{0mm}{5mm}
$p p\to  H \to \tau^+ \tau^-$ & 1.06 & 1.6 & Comb. & 100-150 &  \cite{ATLAS:taus,CMS:taus}
\\ [0.3em]
\hline
\rule{0mm}{5mm}
$VH, H \to b \bar{b}$ & $-$  & 1.1 & CMS $\times$ 2 & 110-135 &  \cite{CMS:VHtobb}
\\ [0.3em]
\hline
\rule{0mm}{5mm}
$qqH, H \to \tau^+ \tau^-$ & 1  & $-$ & ATLAS $\times$ 2 & 110-130 &  \cite{Atlas:HiggsNote}
\\ [0.3em]
\hline
\end{tabular}
\end{center}
\caption{List of LHC channels used in this study. Here, H stands for either $h_1$ or $h_2$. The production mechanisms considered in $p p$ are gluon-fusion, vector boson fusion, associated production with $Z,W, t \bar{t}$ and also $b \bar{b} \to H$. The cross sections at the LHC have been taken from ref~\cite{LHCHiggsCrossSectionWorkingGroup:2011ti}. See the main text for details. 
}\label{tab:LHC_chan}
\end{table}
In our analysis we include the current LHC data and future projections for the search channels listed in table~\ref{tab:LHC_chan}. All of the searches use the most up-to-date LHC data with a total integrated luminosity between 1.04 - 2.28 fb$^{-1}$ (depending on the search channel), except for the  $qqH, H \to \tau^+ \tau^-$ channel, for which we use the MonteCarlo 2010 sample \cite{Atlas:HiggsNote}, which provides a projection of the expected sensitivity of the current data-sample, since no LHC collaboration has presented yet updated data in this search channel. In the case of the associated production with a vector boson, with the Higgs decaying into bottom pairs, the current analysis was done using a cut based procedure that is able to exclude a Higgs boson with a rate of around 20 times the SM case \cite{Atlas:VHtobb}. The MC 2010 analysis was performed by taking advantage of boosted $b \bar{b}$ pairs \cite{Butterworth:2008iy}, and the expected exclusion with 1 fb$^{-1}$ of data for this case is around 6 times the SM \cite{Atlas:HiggsNote}. We note that these channels are not able to probe points in our model, but, for the sake of completeness, we include them in our analysis.

We combine the results from ATLAS and CMS in the channels where both collaborations have presented data, following the prescription detailed in Refs.~\cite{Draper:2009fh,Draper:2009au} (see below). The \emph{current exclusion} is obtained by using the observed limits reported by ATLAS and CMS. For channels where only one of the collaborations has presented data, we will base our current exclusion on that analysis. In the same case, we compute the \emph{future projections} by doubling the expected result in an attempt to mimic the combination of both experiments and scaling the result by the expected total integrated luminosity using the prescription detailed below. Since the reach for SM-like Higgs bosons at CMS and ATLAS is similar, this approximation is expected to be reasonably accurate. Except in the $ZZ$ channel, all of the others involve, for a particular mass range, one definite final state. For the $ZZ$ we also combine in quadrature the results for the a) four leptons, b) two leptons two quarks and c) two leptons plus two neutrinos, and d) two leptons plus two taus (CMS only) final state searches
\begin{figure}[ht]
\begin{center}
\begin{minipage}[b]{0.45\linewidth}
\begin{center}
\includegraphics[width=1\textwidth]{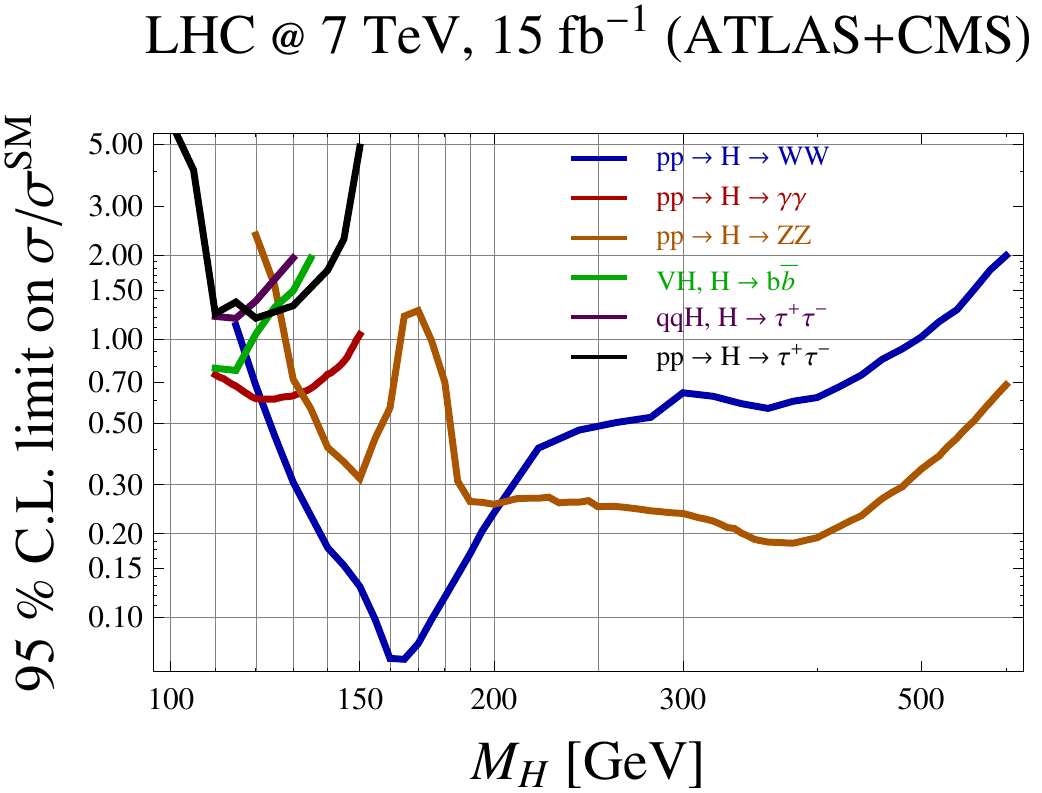}
\newline
(a)
\end{center}
\end{minipage}
\hspace{0.5cm}
\begin{minipage}[b]{0.45\linewidth}
\begin{center}
\includegraphics[width=1\textwidth]{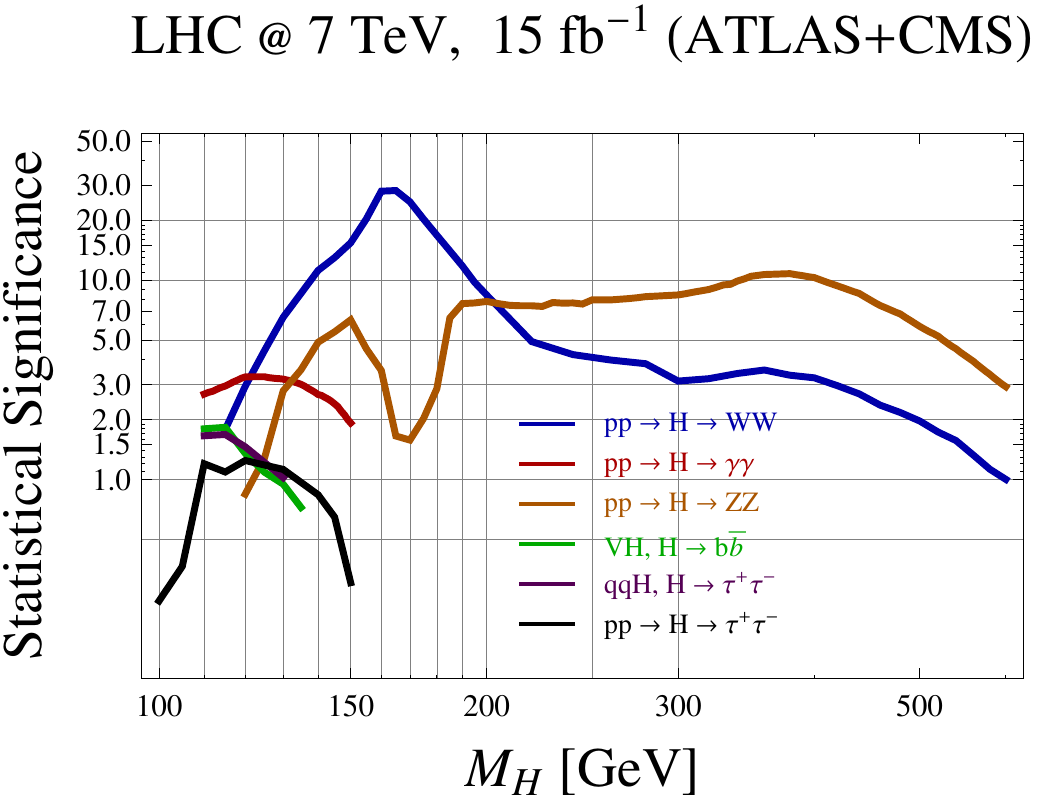}
\newline
(b)
\end{center}
\end{minipage}
\caption{ LHC reach (a) and statistical significances (b) for the SM Higgs boson H
with $15~\rm{fb}^{-1}$, combining both experiments.  The color coding
is as follows: WW (blue), ZZ (orange), $\gamma \gamma$ (red), $\tau
\tau$ (black) , $VH, H \to b \bar{b} $ (green) and $qqH, H \to \tau^+ \tau^-$ (purple).
}\label{fig:lhc_reach}
\end{center}
\end{figure}

Here we briefly review the procedure used to combine the experimental information. For each relevant channel we compute the following quantity:
\be\label{eq:Qdef}
Q ({\cal L}_0)=\frac{R_{\rm{mod}}}{R_{\rm{exp}} ({\cal L}_{0})} \, ,
\ee
where $R_{\rm{mod}}$ is the rate in this particular channel coming from our model, $R_{\rm{exp}} $ is the exclusion limit at the 95\% C.L., at a reference total integrated luminosity ${\cal L}_{0}$. Eq.~(\ref{eq:Qdef}) is exactly the same definition used by HiggsBounds in order to set the 2$\sigma$ exclusions: if $Q > 1$ the point is excluded at the 95\% C.L.. We compute $R_{\rm{exp}} $ combining  the results for each LHC experiment in inverse quadrature (see Refs.~\cite{Draper:2009fh,Draper:2009au}). In ref~\cite{Carena:2011fc} this procedure was found to be more conservative than the combination performed by the ATLAS collaboration by 10-20 \%
\footnote{While the first version of this manuscript was under consideration, ATLAS and CMS presented the combination of their datasets in \cite{Comb:HCP}. We have compared their results against our naive combinations, finding that the expected values differ by at most 10\%, while for the observed values the discrepancy ranges from 30 to 50 \%, but in those cases our naive combination turns out to be a conservative. A similar comparison is also shown in figure~2 of ref~\cite{Boudjema:2011xf}, where the SM combination is confronted against the experimental result, also finding a similar accuracy.}.
While the quantity $R_{\rm{mod}}$ is a number that does not change, $R_{\rm{exp}}$ scales with the luminosity as ${\cal L}^{-1/2}$. Thus, defining $R_{\rm{exp}} ({\cal L}_0) = R_0$ and $Q_0 = R_{\rm{mod}} / R_{0}$, one has that
\be
Q ({\cal L}_1) = \frac{R_{\rm{mod}}}{R_0} \sqrt{ \frac{{\cal L}_1}{{\cal L}_0} } = Q_0 \sqrt{ \frac{{\cal L}_1}{{\cal L}_0} }   \, .
\ee
In order to derive these equations, one is neglecting all systematic effects and also assumes that in each particular channel $B \gg S \gg 1$ holds, where $B$ ($S$) are the number of background (signal) events for a particular channel.
With these simplifications the expected statistical significance $\sigma$ turns out to be $\sigma \approx 2~Q$.

As an illustration, we present in figure~\ref{fig:lhc_reach} the expected reach at the LHC and the statistical significance for the SM Higgs as a function of the Higgs mass in the channels described in table~\ref{tab:LHC_chan}, assuming a total integrated luminosity of 15 fb$^{-1}$, which corresponds to the total integrated luminosity that can be collected by the end of 2012 if the instantaneous luminosity is kept at the current rate.

As one can see, the $WW$ channel is setting the most stringent exclusion in the $120-200$ GeV range. For larger masses, the $H \to ZZ$ channel takes the leading role. For masses below $120$ GeV one enters into the problematic range, where the $WW$ channel becomes ineffective, and the diphoton requires ${\cal O} (10~\rm{fb}^{-1}$) integrated luminosity in order to probe the SM Higgs. Moreover, as mentioned in Section~\ref{subsec:minHiggspot}, the suppression factor for one of the Higgs bosons is at most 1/2, unless there is a significant \emph{non-standard} branching ratio. This means that in the case of maximal mixing without significant extra-SM decay modes, at least one Higgs boson can be tested at the 2 (5)$~\sigma$ level if its mass lies in the 125-550 (140-190) GeV range.

%
\section{Numerical analysis: results}
\label{sec:results}
\begin{figure}[tbp]
\begin{center}
\begin{minipage}[b]{0.45\linewidth}
\begin{center}
\includegraphics[width=1\textwidth]{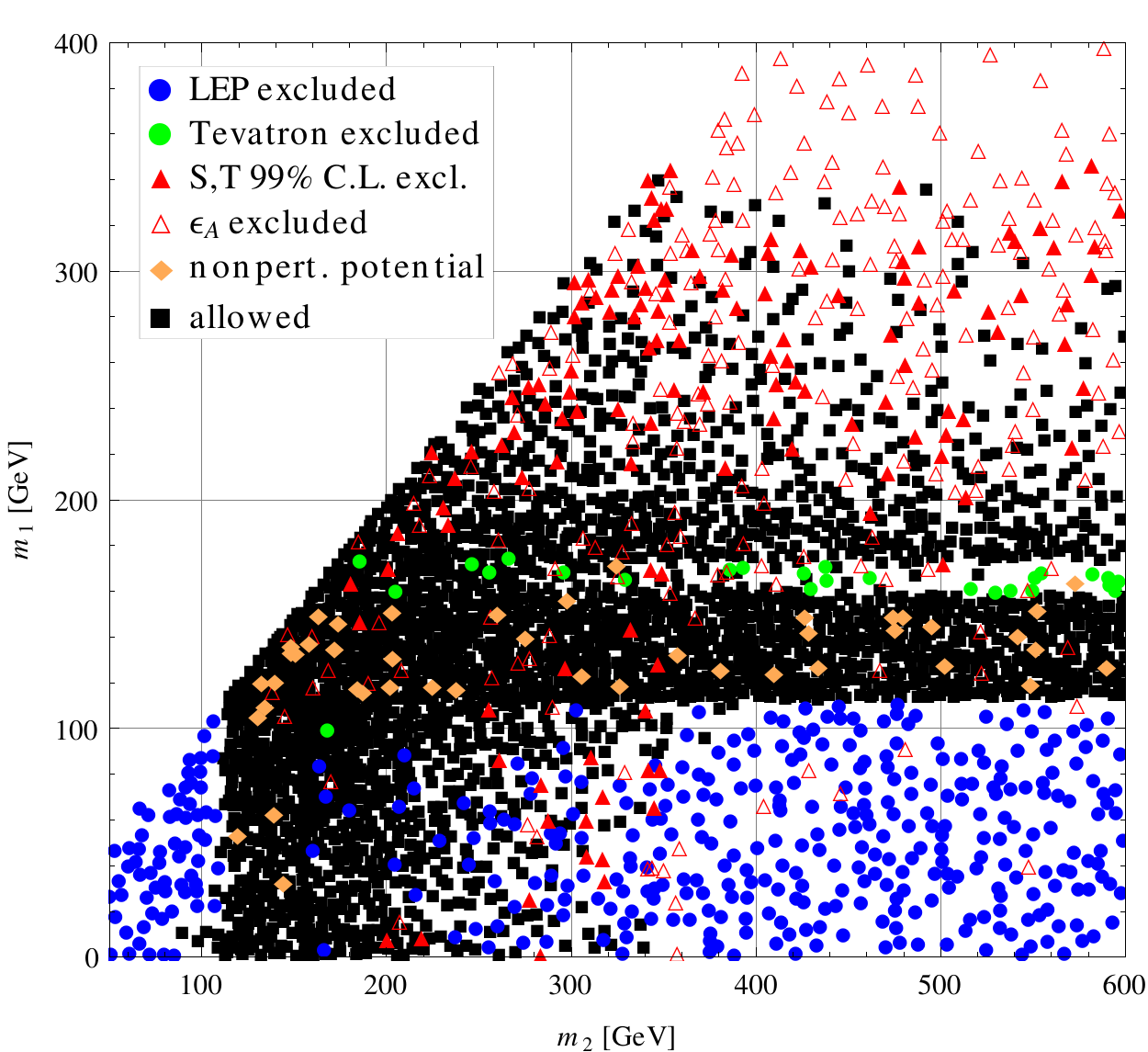}
\newline
(a)
\end{center}
\end{minipage}
\hspace{0.5cm}
\begin{minipage}[b]{0.45\linewidth}
\begin{center}
\includegraphics[width=1\textwidth]{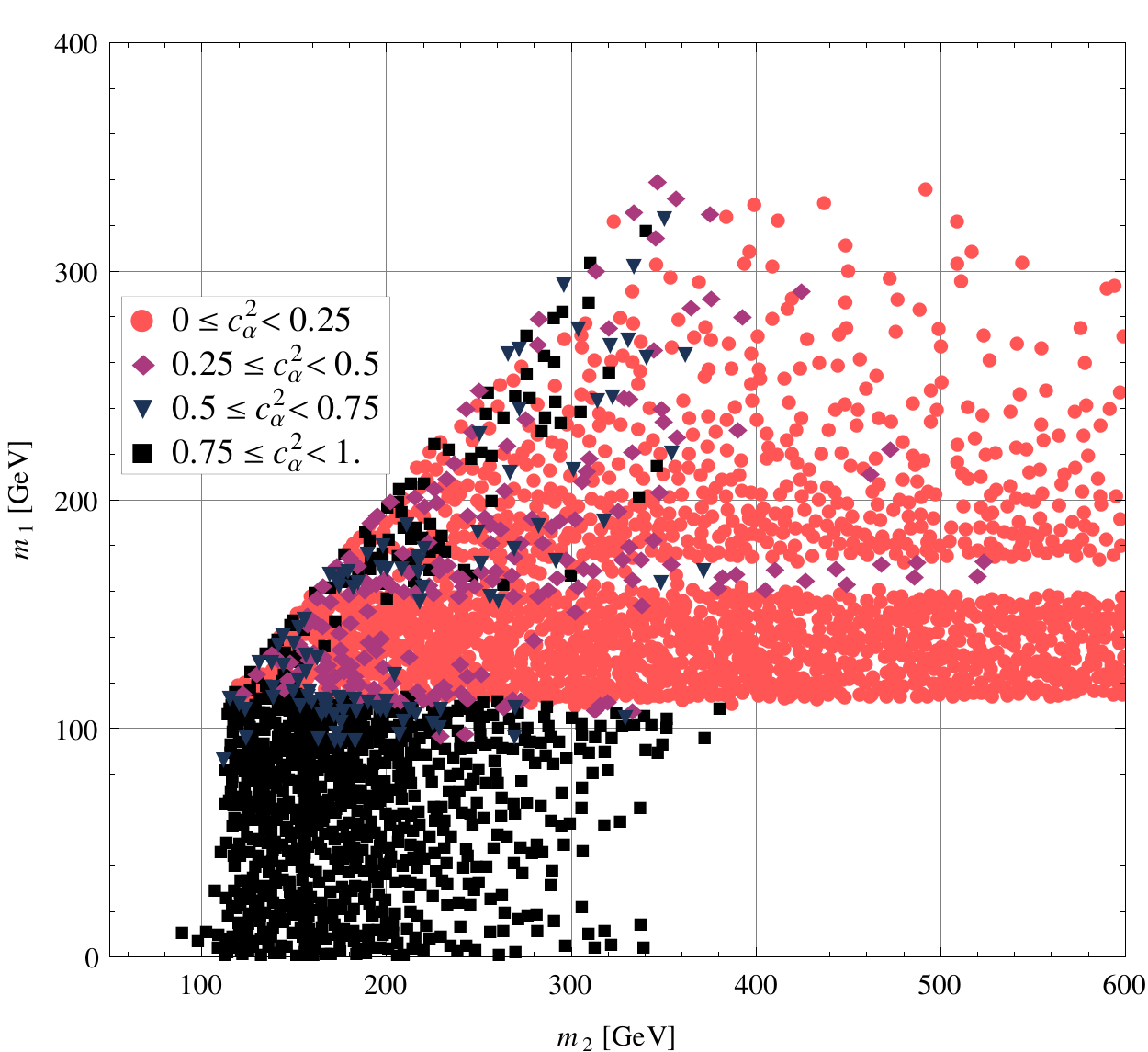}
\newline
(b)
\end{center}
\end{minipage}
\caption{ Scanned points in the $m_{1}  -m_{2}$ plane. In the left panel (a) we show a representative sample of excluded along with allowed points, while in the right panel (b) we show only allowed points. In the left panel, we show the points excluded by LEP (blue), Tevatron (green), S- and T-parameters (red, filled), $\epsilon_A$ (red, empty) and perturbativity constraints (orange). The black points are allowed by all current collider data. In the right panel the color coding varies according to the value of $c_\alpha^2$: red $(0 \le c_{\alpha}^2 < 0.25)$, magenta $(0.25 \le c_{\alpha}^2 < 0.5)$, blue $(0.5 \le c_{\alpha}^2 < 0.75)$ and black $(0.75 \le c_{\alpha}^2 \le 1)$. }
\label{fig:fig1}
\end{center}
\end{figure}
In this section we present the results of the parameter scan as defined in Section~\ref{sec:scans}. 
In the left panel of figure~\ref{fig:fig1} we study the impact of each experimental or theoretical bound on the parameter space of the model in the $m_1 - m_2 $ plane. We plot points that are excluded by LEP (blue), Tevatron (green), S and T parameter (red, filled), $\epsilon_A$ (red, empty) and the requirement of perturbativity of the potential parameters (orange). 
It is clearly visible how the direct searches of LEP (blue) and Tevatron (green) constrain the region $m_1 < 114.4~\rm{GeV}$ and $m_i \approx 160-170~\rm{GeV}$. The indirect bounds via the $S$ and $T$ parameters (red, filled) and the constraints on kinetic mixing of the neutral massive gauge bosons (red, empty) mostly affect regions where one or both Higgs bosons are heavier than 155 GeV.
As can be seen from equation (\ref{eq:lamSM}), the perturbativity requirement $\lambda < 1$ places an upper bound $m_1 < \sqrt{2} v \approx 350$ GeV on the mass of the lighter Higgs boson (orange). However, points in that region tend to be excluded for other reasons before so that its most important effect is to prevent the decay width of either Higgs boson into $Z'Z'$ from becoming large enough to invalidate the narrow width approximation.

The black points evade all of the above constraints and are the focus of the study at hand. This subset is shown in the right panel of figure~\ref{fig:fig1}, where we have colored the points according to the rescaled squared coupling of $h_2$ to Standard Model particles, namely, to the particular value of $c_{\alpha}^2$: red $(0 \le c_{\alpha}^2 < 0.25)$, magenta $(0.25 \le c_{\alpha}^2 < 0.5)$, blue $(0.5 \le c_{\alpha}^2 < 0.75)$ and black $(0.75 \le c_{\alpha}^2 \le 1)$. 

We focus first on the region where the mass of the ligher Higgs state $h_1$ is below the LEP limit of 114.4 GeV. Here, the $h_1$ coupling to Standard Model particles has to be significantly suppressed to avoid direct detection\footnote{The parameter space points where $m_1 < 12 $ GeV should be taken with care, since the LEP search in the $h_1 Z, \, h_1 \to b \bar{b}$ is cut-off at this value \cite{Schael:2006cr} and there are other low energy experiments that can probe this mass range more efficiently than the searches included in this study (see also \cite{arXiv:1006.1151} for an analysis of the LHC reach.)
 . The detailed analysis of this region is outside the scope of the present work. }. 
The heavier state $h_2$ can be lighter than 114.4 GeV at the same time if $h_2 \rightarrow h_1 h_1 \rightarrow \text{SM}$ is by far the dominant decay channel, in which case it can evade the constraints coming from the LEP searches for invisible decays of a Higgs-like scalar\footnote{Recent studies based on jet-substructure techniques show that the $h_2 \to h_1 h_1 \to 4j$ final states can be tested at the $5 \sigma$ level at the 14 TeV LHC with ${\cal O} (10-100~\rm{fb}^{-1}$ of data \cite{arXiv:1006.1650,arXiv:1012.1316,arXiv:1102.0704}, depending on the model under consideration, and on $m_1$ and $m_2$. Due to these reasons we decide not to include those analysis in the present work. }. The decay $h_i \rightarrow Z Z'$ is subdominant (branching fraction below 1\%) in this region as well as in the whole parameter space.

When $m_1 \lesssim 114~\textrm{GeV} \lesssim m_2 \lesssim 155~\rm{GeV}$, the heavy Higgs state $h_2$ behaves largely as the Standard Model Higgs, except for possible non-standard decays. In turn, if $m_2$ is above 155 GeV, fine-tuning between the S and T parameter contributions of the extended bosonic sectors is needed in order to evade the constraints from electroweak precision tests. It is necessary that $m_{Z'} < m_Z$, and for a given point in the $m_1$-$m_2$-plane the allowed range for $m_{Z'}$ becomes smaller, the larger $m_2$ is.

When $m_1$ is above 114 GeV, there are no direct constraints on the mixing angle. The indirect constraints via the $S$ and $T$ parameter force the heavy Higgs to mostly decouple from the Standard Model for masses $m_2 \gtrsim 150$ GeV. Where the decay $h_2 \rightarrow h_1 h_1$ is possible, the corresponding branching fraction is always smaller than $0.5$. $BR(h_2 \rightarrow Z' Z')$ can take any value, whereas $BR(h_1 \rightarrow Z' Z')$ decreases with growing $m_2$. The region where both Higgs masses are larger than 155 GeV is, again, the result of $m_{Z'} < m_Z$.
\begin{figure}[!htbp] \label{fig:2}
\begin{center}
\begin{minipage}[b]{0.45\linewidth}
\begin{center}
\includegraphics[width=1\textwidth]{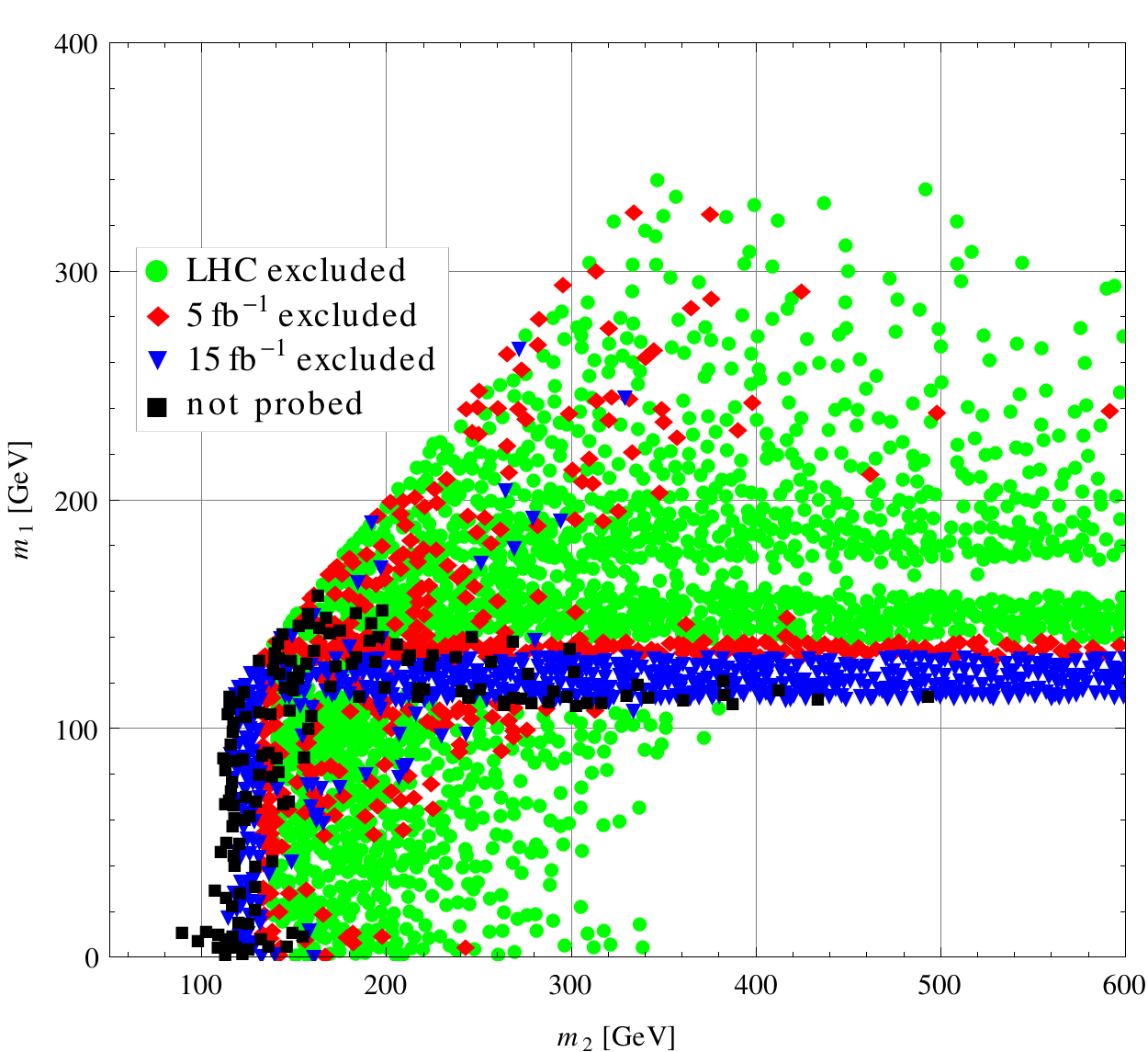}
\newline
(a)
\end{center}
\end{minipage}
\hspace{0.5cm}
\begin{minipage}[b]{0.45\linewidth}
\begin{center}
\includegraphics[width=1\textwidth]{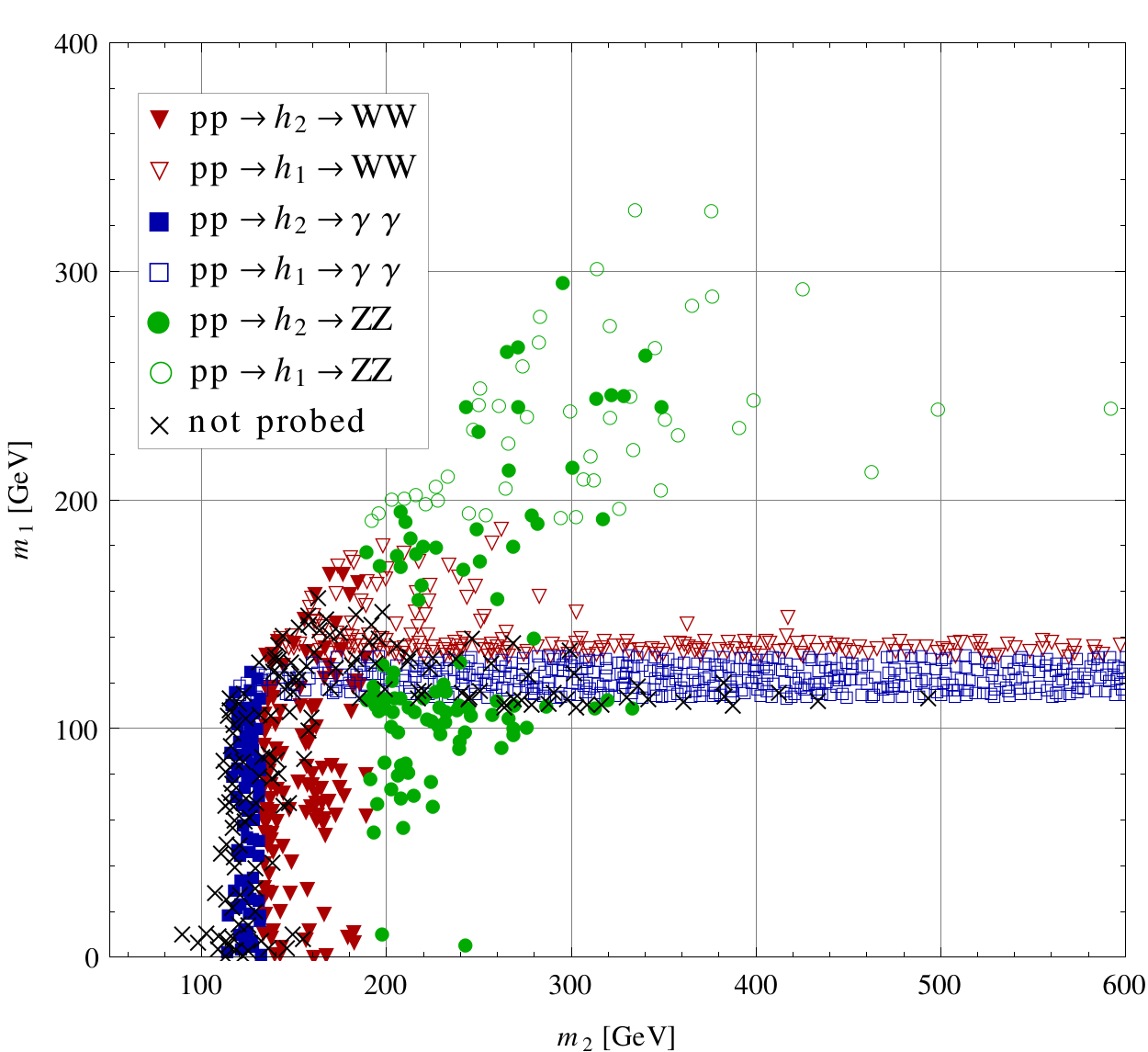}
\newline
(b)
\end{center}
\end{minipage}
\caption{ (a) LHC $2\sigma$ exclusions for different scenarios and (b) most sensitive channels in the $m_{1}  -m_2$ plane. In the left panel, we show the points that have been excluded by the LHC with the current dataset (green), and also those that can be excluded after collecting $5 ~\text{fb}^{-1}$ (red) and $15 ~\text{fb}^{-1}$ (blue) of data. The black points will still be allowed by LHC data. 
In the right panel we remove the points excluded by the LHC today. Color coding varies according to the most sensitive channel to that particular point and filled (empty) shapes correspond to $h_2$ ($h_1$). The $pp \to h_i \to WW$ decay mode is shown in red, the diphoton channel in blue and the $ZZ$ channel in green.  The black points are not sensitive to the LHC search channels under consideration.}\label{fig:fig2}
\end{center}
\end{figure}

Let us now study how the LHC experiments will probe the parameter space with their main search channels for the Standard Model Higgs boson. Even though production rates and decay widths are never enhanced in this model, the parameter space is already probed efficiently. As it can be seen on the left panel of figure~\ref{fig:fig2}, the current dataset (1.04-2.28 fb$^{-1}$), shown in green, is able to exclude a vast majority of points in the region of $m_1 >$  140 GeV, mostly due to the $h_1 \to WW / ZZ$ channels. If $m_1 < 114$ GeV, $h_2 \to WW / ZZ$ is probing values of $m_2$  above 135 GeV range.
With 5 $\text{fb}^{-1}$ (red points) one can exclude almost the complete region $m_1 > 130$ GeV except in the case that $h_1$ has small couplings to SM particles and at the same time a large branching fraction into $Z'Z'$ or other invisible particles.
With 15 fb$^{-1}$ of data (expected in 2012) the diphoton channel will start to probe points in the 110-130 GeV range (blue points in the left plot). The region where both $h_2$ and $h_1$ are below the LEP limit (black points), and where the main decay mode for $h_2$ is into $h_1 h_1$, can not be tested with the channels used in this study, since the LHC collaborations have not presented dedicated searches for this kind of decays (one would typically look into $b \bar{b} b \bar{b}$, $\tau^+ \tau^- b \bar{b}$ or even $\tau^+ \tau^- \tau^+ \tau^-$).  

\begin{figure}[!htbp]
\begin{center}
\begin{minipage}[b]{0.45\linewidth}
\begin{center}
\includegraphics[width=1\textwidth]{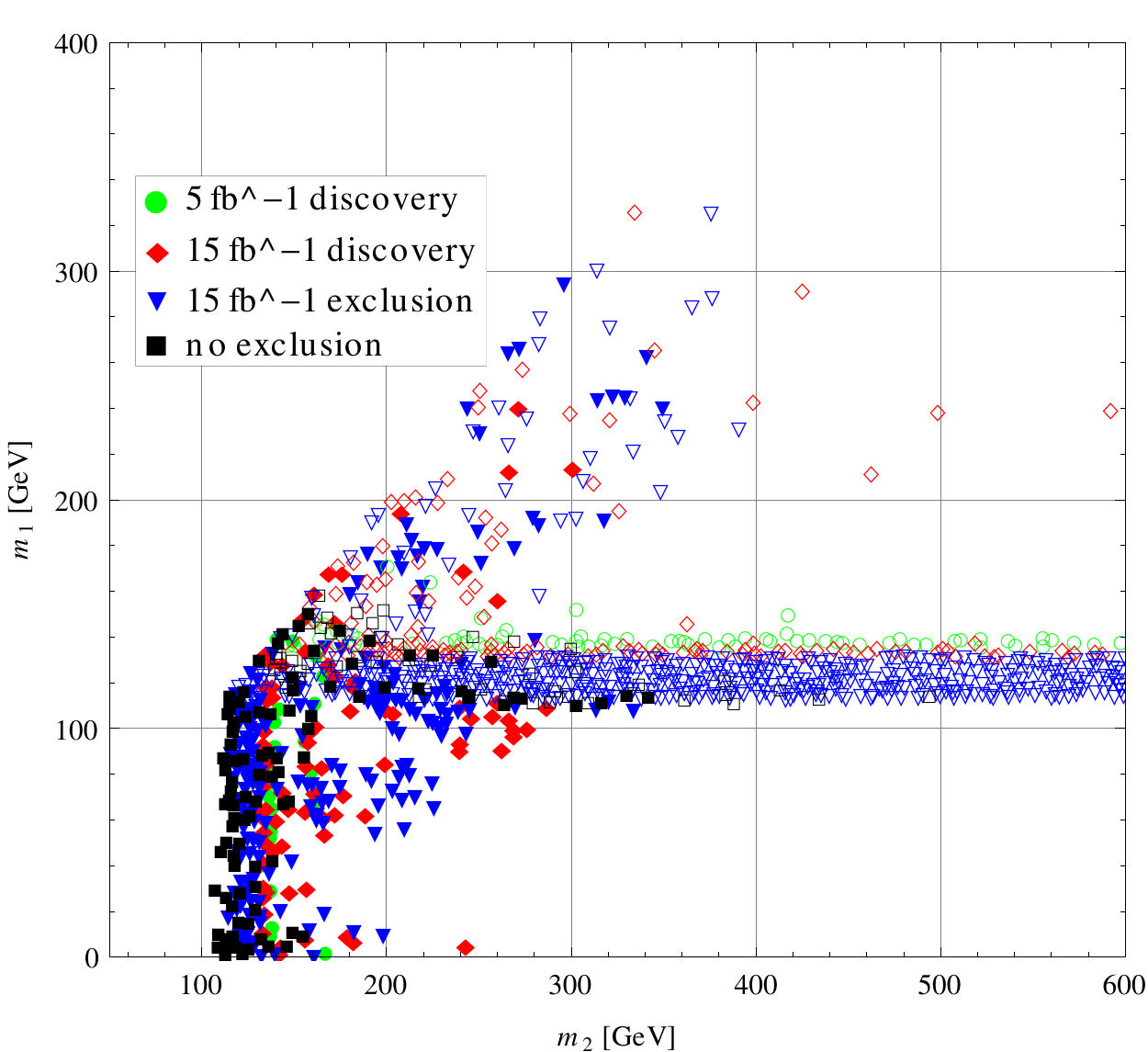}
\newline
(a)
\end{center}
\end{minipage}
\hspace{0.5cm}
\begin{minipage}[b]{0.45\linewidth}
\begin{center}
\includegraphics[width=1\textwidth]{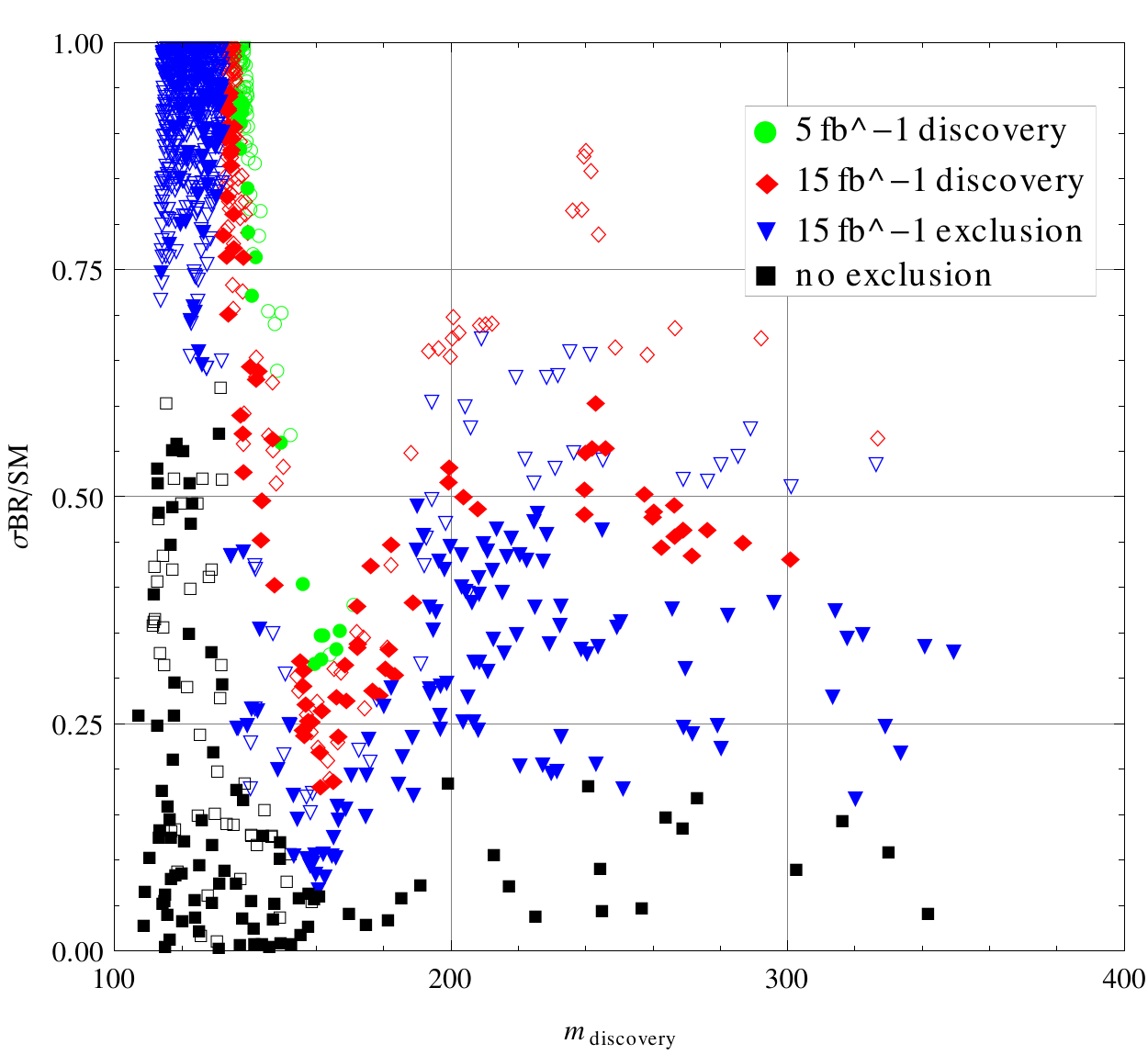}
\newline
(b)
\end{center}
\end{minipage}
\caption{ Left panel: $5 \sigma$ discovery potential with $5~\textrm{fb}^{-1}$ (green), $15~\textrm{fb}^{-1}$ (red), exclusion at $15~\textrm{fb}^{-1}$ (blue) and parameter space points outside of early LHC reach (black). Filled (empty) shapes correspond to the discovered or excluded particle being $h_2$ ($h_1$). Right panel: Rate suppression factor (\ref{eq:ratesup}) at the LHC as a function of the mass of the Higgs more likely to be detected first.}\label{fig:fig3}
\end{center}
\end{figure}

In figure~\ref{fig:fig3} we show the $5\sigma$ discovery potential of the model in the $m_1-m_2$ plane (left panel) and the rate suppression factor (\ref{eq:ratesup}) for the most sensitive search channel as a function of the Higgs mass which is more sensitive for exclusion/discovery at the LHC (right panel).

From  the left panel we see that a discovery with 15 fb$^{-1}$ is only possible if either mass is larger than about 130 GeV. We have explicitly checked that the case where LHC discovers both Higgs states in the early run is only possible if the masses are in the range $130 \lesssim m_1 \lesssim 170$ GeV and $130 \lesssim m_2 \lesssim 260$ GeV and the mixing between the two states is sizable.

When a Higgs scalar with a lowered rate $\sigma \cdot BR$ than the Standard Model expectation 
is detected at the LHC it is a priori impossible to decide which of the two mass eigenstates has been discovered using above searches. Furthermore, more involved studies are needed to determine whether the origin of the rate suppression is the mixing between the states or decays invisible to the specific search channel, e.g. $h_i \rightarrow Z' Z'$ or $h_2 \rightarrow h_1 h_1$. From the right panel we see that after collecting $15~\rm{fb}^{-1}$ of data, one can exclude a rate which is 0.6 (0.05) times the SM rate for $m_h \sim 130$ $(160)$ GeV; for $m_h >$ 200 GeV this value is 0.2. If a Higgs in the mass range of $180 \lesssim m_{\text{Discovery}} \lesssim 300$ GeV is discovered with a moderate rate suppression factor $\sigma\cdot BR / SM \approx 0.7$, it is likely that the detected particle is the lighter mass eigenstate $h_1$. Furthermore, the early discovery of a Higgs state with a mass larger than 155 GeV points toward $m_Z'< m_Z$. This is because it is likely that the first Higgs to be discovered is the one that couples more strongly to the Standard Model. If it is heavier than 155 GeV, a $Z'$ gauge boson with specific properties is needed to reconcile such a high Higgs mass with electroweak precision data.
\section{Conclusions}
\label{sec:conclu}
%
%
In this work we have studied in detail the constraints on dark Higgs models at the 7 TeV LHC. In the scenario under consideration the usual SM Higgs boson (i.e the one responsible for electroweak symmetry breaking) mixes with a complex singlet that breaks an extra $U(1)_D$ gauge symmetry, which in turn mixes with the SM hypercharge $U(1)_Y$. 
The free parameters of this model are the masses of the two Higgs bosons $m_1$ and $m_2$, the cosine of the mixing angle $c_\alpha$ between them, the mass of the additional gauge boson $m_{Z'}$ with the kinetic mixing parameter $\epsilon_A$ and the coupling strength $g_D$. A parameter scan was performed and the effect of theoretical and experimental bounds from direct searches at LEP and Tevatron and electroweak precision data was studied.

The current LHC published analyses (with luminosities between 1.04 and 2.28$~\rm{fb}^{-1}$) are able to exclude masses above 140 GeV, unless there is a significant mixing between the two mass eigenstates, while with $5~\rm{fb}^{-1}$ most of the points with the lightest Higgs mass above 130 GeV are excluded. Furthermore, we have found that the 7 TeV LHC with $15~\rm{fb}^{-1}$ of total integrated luminosity will be able to probe most of the parameter space at the $2\sigma$ level. The points that evade these constraints correspond to two cases.
The first of them is when  either Higgs boson lies in the $115-130$ GeV range and there is some non-negligible mixing between the two mass eigenstates. For such masses the $WW$ final state is not very useful, and if there is some mixing between $h_1$ and $h_2$ one can loosen the exclusion power of the $\gamma \gamma$ channel, which will rule out that mass range for the SM Higgs boson. However, we would like to point out that, according to projection on the SM Higgs that include combinations of the different low-mass sensitive channels, this region can in principle be accessed during the 7 TeV run.

The second case takes place when the Higgs rates are diminished due to a sizable mixing between the mass eigenstastes, or if $h_2 \to h_1 h_1$ is kinematically open. For this region of parameter space one should perform a dedicated search of $h_2 \to h_1 h_1$ by looking at final states like $b \bar{b} b \bar{b}$, $\tau^+ \tau^- b \bar{b}$ or even $\tau^+ \tau^- \tau^+ \tau^-$. 

We would like to remark that the effectiveness of these bounds gets looser if there is an important partial width of any Higgs boson into other (not specified in this work) dark sector degrees of freedom (like the dark matter candidate). 
As a consequence, our results can either be interpreted  as valid in a completion of the model where the aforementioned channel is not relevant, or also as the largest exclusion coverage one can get in parameter space.

We have also analyzed the possibility of discovering one or two Higgs bosons. We have found that, with 15 fb$^{-1}$ one can discover one of the Higgs bosons of this scenario if their masses are larger than 130 GeV. 
If only one Higgs boson with a rate smaller than the SM Higgs is discovered, it would be impossible to tell \emph{a priori} whether it is $h_1$ or $h_2$. Discovering both Higgs bosons will only happen if $m_2 \in [130,260]~\rm{GeV}$ and $m_1 \in [130,170]~\rm{GeV}$. Such an observation would rule out some other models, like for instance the MSSM, since the lightest neutral Higgs boson can not have a mass well above 135 GeV. 
Finally, if the LHC should not see any Higgs signature after collecting $15~\rm{fb}^{-1}$ of data then one can use that result to constrain the mixing between the two Higgs bosons and/or the invisible width. For the former case, a recent study using the $H \to ZZ \to 4l$ lineshape was presented in ref~\cite{Low:2011kp}.
\acknowledgments{
We would like to thank Thomas Gehrmann for numerous discussions during the project and comments on the manuscript. We also thank Marc Gillioz for proofreading the article. 
This work is supported by the Swiss National Science Foundation (SNF) under contract 200020-138206 and in part by the European Commission through the
``LHCPhenoNet" Initial Training Network PITN-GA-2010-264564.}

Note added: while this paper was in press, the LHC collaborations have shown updated results using a larger dataset (about 5 fb$^{-1}$ of data). Their results hint on a signal of a Higgs boson with a mass of around 125 GeV with rates compatible with the SM. These results leave our conclusions unchanged.
%
%

%
\end{document}